\documentclass[reprint,aps,prl,showpacs,twocolumn,letterpaper]{revtex4}
\usepackage{amssymb,amsfonts,amsmath}
\usepackage{times}

\usepackage{graphicx}

\usepackage{color}

\begin{document}

\title{Superconducting High Pressure Phases Composed of Hydrogen and Iodine}
\author{Andrew Shamp}
\author{Eva Zurek}\email{ezurek@buffalo.edu}
\affiliation{Department of Chemistry, State University of New York at Buffalo, Buffalo, NY 14260-3000, USA}

\begin{abstract}
Evolutionary structure searches predict three new phases of iodine polyhydrides stable under pressure. Insulating $P_1$-H$_5$I, consisting of zigzag chains of (HI)$^{\delta+}$ and H$_2^{\delta-}$ molecules, is stable between 30-90~GPa.  $Cmcm$-H$_2$I and  $P6/mmm$-H$_4$I are found on the 100, 150 and 200~GPa convex hulls. These two phases are good metals, even at 1~atm, because they consist of monoatomic lattices of iodine. At 100~GPa the $T_c$ of H$_2$I and H$_4$I are estimated to be 7.8 and 17.5~K, respectively. The increase in $T_c$ relative to elemental iodine results from a larger $\omega_\text{log}$ from the light mass of hydrogen, and an enhanced $\lambda$ from modes containing H/I and H/H vibrations.
\end{abstract}

\maketitle

We have only begun to uncover the affects of pressure on the stoichiometries, electronic structures and emergent behavior  of condensed matter \cite{Grochala:2007a,Zurek:2014i}. A number of phases with combinations that appear strange from a 1~atm perspective have been theoretically predicted and experimentally synthesized. These include  
hydrides with unique combinations such as LiH$_2$ and LiH$_6$ \cite{Zurek:2009c,Pepin:2015a}, as well as H$_n$S phases \cite{Strobel:2011a}, some of which may have a superconducting transition temperature, $T_c=190$~K at 150~GPa \cite{hns,Cui:2014a,Errea:2015a,Flores:arxiv}, that surpasses those of the oxide-based high-temperature superconductors. Other predicted phases include sub- or polyhydrides containing 
an alkali or alkaline earth metal \cite{Zurek:2011d,Zurek:2012a,Zurek:2009c,Zurek:2011h,Zurek:2012b,Zurek:2012g,Zurek:2012m,Zurek:2012n,Zurek:2013f,Ma:2012f} and hydrides containing a Group 14 element \cite{Gao:2008a,Chen:2008b,Canales:2009a,Canales:2006a,Pickard:2006a,Yao:2010a,Livas:2012a,Kim:2008a,Tse:2007a}. Theory has suggested that some of these hydrides may be superconducting at high temperatures \cite{Ma:2012f}, or contain unusual structural motifs such as the linear symmetric H$_3^-$ anion  -- the simplest example of a three-center-four-electron bond \cite{Zurek:2012g}.

Another example of unique stoichiometries that could potentially be synthesized under pressure are H$_2$Cl and H$_5$Cl \cite{Ma:2015a}. They were predicted to become stable with respect to decomposition into H$_2$ and HCl at 60~GPa, and remain so until at least 300~GPa. All of the H$_n$Cl phases contained zigzag [HCl]$_\infty$ chains reminiscent of those present within HCl. The three-center-two-electron (3c-2e) H$_3^+$ motif, which approached the ideal equilateral triangle configuration by 300~GPa, was incorporated within H$_5$Cl.  How would decreasing the electronegativity of the dopant atom affect the stable crystal lattices and their electronic structures? To answer this question we theoretically studied the iodine polyhydrides, H$_n$I. A further motivation for our work: it has been proposed that pressure can be used to synthesize alloys of metallic hydrogen with the general formula (H$_2$)$_{1-x}$HI$_x$ that are potentially superconducting \cite{Straaten:1986a}. Moreover, HI undergoes a pressure induced insulator to metal transition below 50~GPa \cite{Straaten:1986a}, and iodine becomes metallic at 16~GPa \cite{Drickamer:1963a} and superconducting below 1.2~K at 28~GPa \cite{Shimizu:1994a}.

The isomorphic low temperature phases of HF, HCl and HBr contain planar zigzag chains of hydrogen-bonded molecules held together by weak van der Waals (vdW) forces. HI assumes a planar distorted hydrogen-bonded diamond lattice, but its detailed structure is not known \cite{Ikram:1993a}. Density functional theory (DFT) calculations have been undertaken to study the structural evolution of the heavier hydrogen halides up to 200~GPa \cite{ZhangMa:2010a}. But, because the phases of HI under pressure are unknown we proceeded to find the global minimum structures from 0-200~GPa in 50~GPa intervals using the evolutionary algorithm (EA) \textsc{XtalOpt} \cite{Zurek:2011a} coupled with PBE-DFT \cite{Perdew:1996a} calculations carried out with \textsc{VASP} \cite{Kresse:1993a}. More information about the computational details is provided in the Supplemental Material, SI \cite{HnISI}. The most stable structures that emerged in our EA searches were comprised of 2D segregated layers of H$_2$ and iodine, implying that HI is metastable with respect to these constituents. In addition, we optimized geometries where HI assumed the experimentally determined and theoretically predicted low-temperature structures of the lighter hydrogen halides \cite{ZhangMa:2010a}. All of these were less stable than the H$_2$/I segregated phases. These results suggest that the impurities and side-products observed upon compression of HI above 70~GPa \cite{Straaten:1988a} may have been due to decomposition of the halide. A previous theoretical study found that HBr was unstable with respect to decomposition into  H$_2$ and Br$_2$ between 120-150~GPa \cite{ZhangMa:2010a}. At these pressures bromine is monoatomic, as is iodine, and it should not be unexpected that HI behaves similarly but at lower pressures.
\begin{figure}
\begin{center}
\includegraphics[width=0.9\columnwidth]{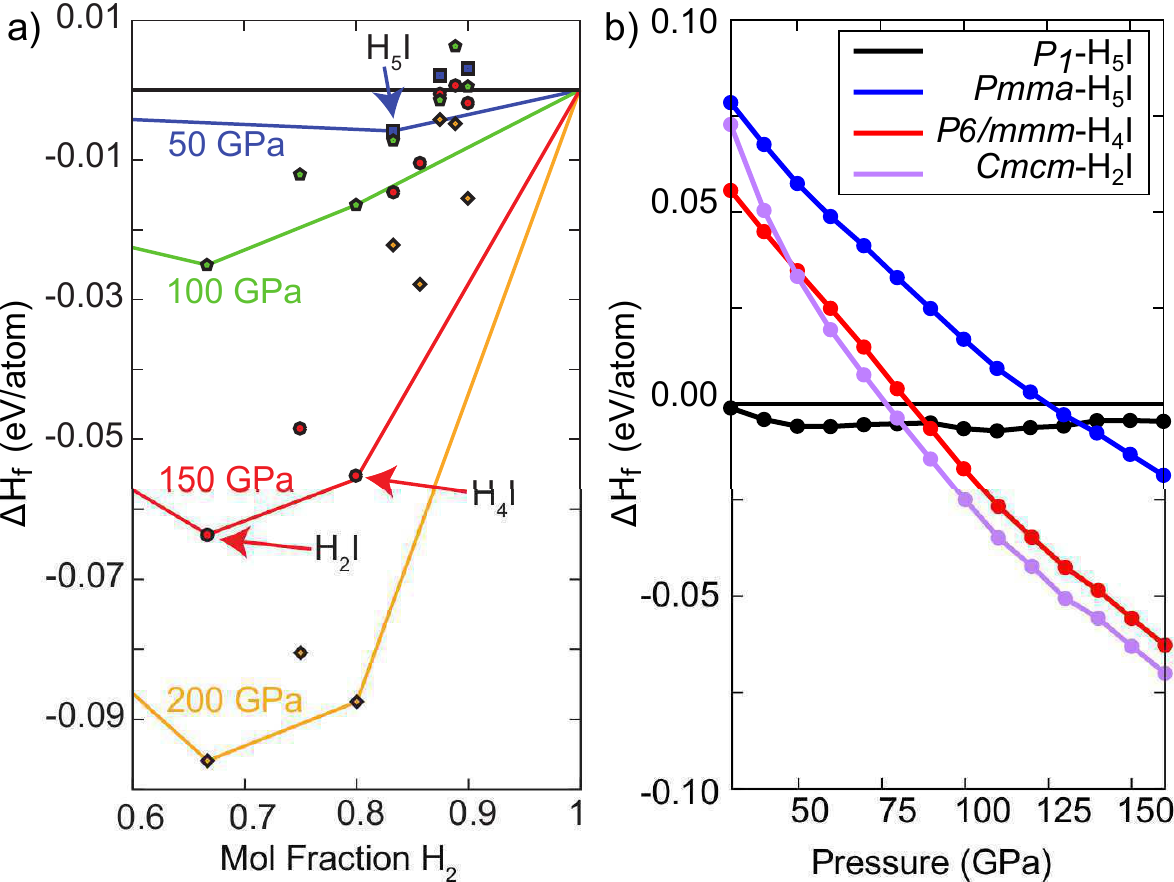}
\end{center}
\caption{(a) The enthalpy of formation, $\Delta H_f$, for the reaction ($\frac{1}{2}$H$_2$)$_n$+I $\rightarrow$ H$_n$I vs.\ the H$_2$ composition as a function of pressure. For H$_2$ we used the enthalpy of the structures from Ref.\ \cite{Pickard:2007a} and for iodine the structure sequence $Cmca\rightarrow Immm\rightarrow I4/mmm\rightarrow Fm\bar{3}m$ \cite{Shimomura:1978a,Takemura:1980a,Fujii:1986a,Fujii:1987a}. The computed transition pressures of 18, 42, and 55~GPa are in good agreement with the experimental ones, 21, 43, 55~GPa (neglecting the incommensurately modulated phase \cite{Kenichi:2003a}). $Fm\bar{3}m$ iodine persists until at least 275~GPa \cite{Fujii:1987a}. (b) $\Delta H_f$ vs.\ pressure of select H$_n$I phases. $\Delta H_f$(H$_5$I) becomes negative for $P>30$~GPa. Above 90~GPa H$_2$I has the most negative $\Delta H_f$.
\label{fig:IHnTieline}}
\end{figure}

EA runs were carried out to find the most stable crystal lattices of H$_n$I ($n$~=~2-9) at 50, 100, 150 and 200~GPa. The structural motifs present in the hydrogen-rich iodine phases that emerged from our searches fall into one of two categories: I$^{\delta+}$, H$^{\delta+}$, and H$_2^{\delta-}$ molecules as found in H$_5$I, or I$^{\delta+}$ and H$_2^{\delta-}$ molecules as found in  H$_2$I and H$_4$I. In Ref.\ \cite{Ma:2015a} it was shown that density functionals approximating the effects of vdW interactions, and zero-point-energy contributions to the enthalpies both altered the transition pressures between the most stable H$_n$Cl phases slightly, but they did not affect the identify of the phases nor their stability. For this reason we also expect them to have a negligible effect on our results.

Fig.\ \ref{fig:IHnTieline} shows the $\Delta H_f$ of the predicted phases with respect to the elemental solids. The convex hull is defined as the set of line segments below which no other points lie, and the phases whose $\Delta H_f$ comprise the hull are thermodynamically stable with respect to decomposition into other polyhydrides and/or solid H$_2$ and iodine. At 50~GPa H$_5$I is the only species on the hull. This stoichiometry becomes stable with respect to decomposition into H$_2$ and I at $\sim$30~GPa and it has the most negative $\Delta H_f$ of any of the structures examined until 90~GPa. At this pressure the H$_2$I stoichiometry becomes the lowest point on the hull and remains so until at least 200~GPa. H$_4$I also comprises the 100, 150 and 200~GPa hulls. At 200~GPa the $\Delta H_f$ of all of the phases continues to decrease. Even though some of the stoichiometries predicted to be stable in our calculations were also found in the hydrogen rich H/Cl phase diagram under pressure \cite{Ma:2015a}, the structures adopted and their properties are distinct, highlighting the importance of the electronegativity  (2.66 for I and 3.16 for Cl on the Pauling scale) and radius (1.98~\AA{} for I and 1.75~\AA{} for Cl) of the halogen atom on the emerging structures and their properties. Importantly, whereas the H$_n$Cl structures were not superconducting \cite{Ma:2015a}, we predict that electricity may pass without resistance through H$_2$I and H$_4$I below $\sim$7.8 and 17.5~K at 100~GPa, respectively.

\begin{figure}
\begin{center}
\includegraphics[width=1.0\columnwidth]{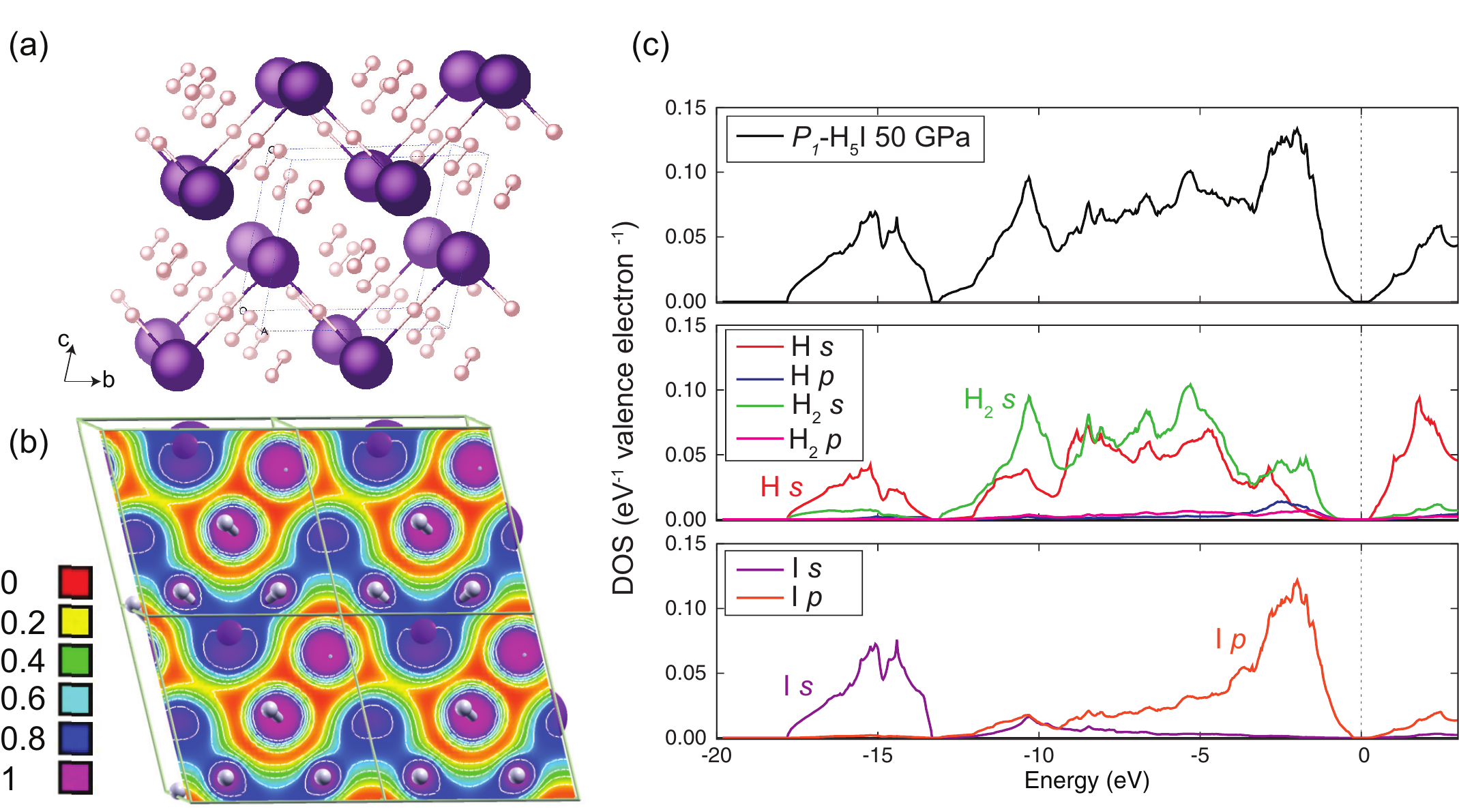}
\end{center}
\caption{(a) Supercell of $P_1$-H$_5$I at 50~GPa. Hydrogen atoms are white, iodines are purple. The zigzag HI chains are a common motif in compressed hydrogen halides. (b) Contour plot of the ELF map of $P_1$-H$_5$I shown in a plane parallel to the $b$-axis. (c) Total and projected DOS of $P_1$-H$_5$I at 50 GPa. $E_F$ has been set to zero.
\label{fig:H5Ifigure2}}
\end{figure}

The most stable H$_5$I phase possessed $P_1$ symmetry up to 90~GPa, and it assumed $Pmma$ symmetry at higher pressures. Above 150~GPa another phase, containing only molecular hydrogen, became preferred, but it did not fall on the convex hull. Phonon calculations at 50~GPa verified the dynamic stability of $P_1$-H$_5$I. Just like the H$_n$Cl phases predicted in Ref.\ \cite{Ma:2015a}, H$_5$I contained zigzag [HX]$_\infty$ (X=Cl, I) chains that resembled those found in HCl, HBr, and HI at low temperatures \cite{ZhangMa:2010a}. These chains lay parallel to the crystallographic $b$-axis in $P_1$-H$_5$I, as illustrated in Fig.\ \ref{fig:H5Ifigure2}. At 50~GPa the H-I distances along the chains were nearly equivalent, 1.795 and 1.788~\AA{}, indicative of pressure induced multi-center bonding, which has also been observed in HX (X=F, Cl, Br) \cite{ZhangMa:2010a} and H$_2$Cl \cite{Ma:2015a}. The H-I-H and I-H-I angles measured 85.7-85.8$^\circ$ and 179.4-179.8$^\circ$. The chains arranged to form sheets that were surrounded by layers of H$_2$ units in which the H-H bond measured 0.747~\AA{}, which is nearly identical to what we calculate for the isolated gas phase molecule at 1~atm, 0.750~\AA{}. Actually, six H$_2$ molecules surround half of the hydrogen atoms in the [HI]$_\infty$ chains in a ``side-on'' fashion. In our previous studies we found a number of structures where H$_2$ encircled the electropositive metal in this same manner, for example in MgH$_{12}$ and MgH$_{16}$ \cite{Zurek:2012m}. A Bader charge analysis revealed that the charge on the hydrogen and iodine atoms in the H-I chains are +0.03 and +0.12, respectively. This can be compared to -0.01 and +0.01 for the hydrogen and iodine atoms in a hypothetical HI phase at 50~GPa whose $Cmc2_1$ symmetry was chosen because HF, HCl and HBr assume this structure at ambient pressure. All of the hydrogen atoms in the H$_2$ molecules were found to bear a slight negative charge such that the average is (H$^{-0.04}$)$_2$. The distance between the hydrogen atoms comprising H$_2$ and the hydrogen atoms in the H-I chains ranged from 1.950-2.022~\AA{} and the closest iodine atom was at least 2.273~\AA{} away. This suggests that the H$_2$ molecules do not interact strongly with any of the atoms comprising the [HI]$_\infty$ zigzag chains. Indeed, the H$_2$ vibron frequency ranged from $\sim$4150-4250~cm$^{-1}$, which compares well with the 4161~cm$^{-1}$ observed for gaseous H$_2$ at 1~atm \cite{Stoicheff}. To probe this further, we calculated the electron localization function (ELF) for $P_1$-H$_5$I at 50~GPa, see Fig.\ \ref{fig:H5Ifigure2}(b). The ELF reveals a high tendency of electron pairing within the H$_2^{\delta-}$ units and along the nearly symmetric [HI]$_\infty$ chains. In fact, the ELF along the chains is reminiscent of the one calculated for symmetric HBr at 30~GPa \cite{ZhangMa:2010a}.

$Cc$-H$_5$Cl was found to be a stable structure in the hydrogen rich H/Cl phase diagram between 100-300~GPa \cite{Ma:2015a}. By 300~GPa H$_3^+$, the simplest example of a 3c-2e bond, emerged in this phase. The bond lengths ($\sim$0.89~\AA{}) and angles (60$^\circ$) in the solid approached those in the gas phase species, and signatures of H$_3^+$ vibrations were observed in the phonon DOS between 1800-3400~cm$^{-1}$. We do not see any evidence for the formation of H$_3^+$ within H$_5$I up to at least 200~GPa.

At 50~GPa $P_1$-H$_5$I is insulating with a PBE band gap of $\sim$0.5~eV, see Fig.\ \ref{fig:H5Ifigure2}(b). The I $s$-states overlap primarily with the H$^{\delta+}$ $s$-states, whereas I $p$-character is found in a region containing both H$_2^{\delta-}$ and H$^{\delta+}$ contributions. Metalization occurs as a result of pressure induced broadening of the valence band, primarily of iodine $p$-character, and the conduction band, which exhibits mainly H$^{\delta+}$ $s$-character, by 65~GPa within PBE. But, because GGA functionals underestimate the pressure associated with band gap closure it is likely that this phase would metalize at a pressure outside the stability range of the H$_5$I stoichiometry. For comparison, HCl and HBr were computed to be metallic by 130 and 80~GPa, and the bands at $E_F$ were primarily of halogen $p$-character \cite{ZhangMa:2010a}.

\begin{figure}
\begin{center}
\includegraphics[width=1\columnwidth]{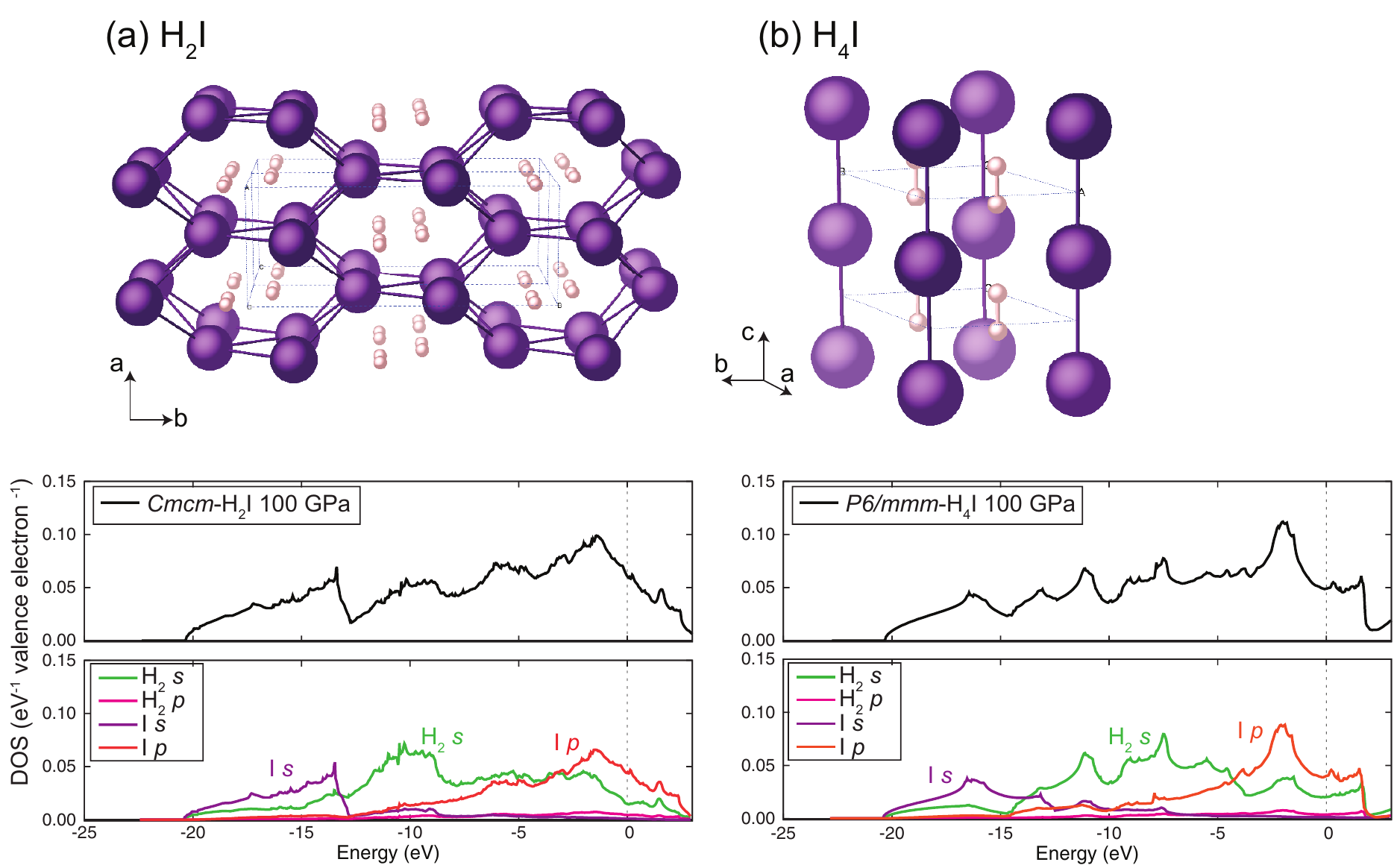}
\end{center}
\caption{Supercells and the total and projected DOS plots of (a) $Cmcm$-H$_2$I and (b) $P6/mmm$-H$_4$I at 100~GPa. 
\label{fig:H2Ifigure1}}
\end{figure}

Between 90-200~GPa a H$_2$I phase with $Cmcm$ symmetry had the most negative $\Delta H_f$, and phonon calculations revealed that it was dynamically stable at 100~GPa. This structure was found to be the most stable H$_2$I arrangement in evolutionary searches performed at 50-200~GPa. Two rows of H$_2$ molecules arranged in a zigzag fashion run through channels formed by the iodine host within this structure, see Fig.\ \ref{fig:H2Ifigure1}(a). At 100~GPa the hydrogen molecules measured 0.796~\AA{} and the distance between them 1.399~\AA{}, as compared with 0.732 and 1.495~\AA{} in solid H$_2$ at this pressure. The I-I contacts measured between 2.896-2.904~\AA{}, which is very close to the  2.899-2.920~\AA{} we compute within $Fm\bar{3}m$ iodine at 100~GPa. The shortest H-I contacts were 2.097~\AA{}, suggesting little interaction between the constituents of the two lattices. $Cmcm$-H$_2$I  bears no resemblance to the H$_2$Cl phase that was predicted to be stable by 100~GPa \cite{Ma:2015a}. High ELF values were calculated between the hydrogens comprising the H$_2^{\delta-}$ molecules. Similar to compressed solid iodine, a large ELF between the iodine atoms was not observed. A Bader analysis showed that charge is transferred from iodine to hydrogen, (H$^{-0.07}$)$_2$ I$^{0.13}$, at 100 GPa. By 200~GPa the H$_2$ bonds within $Cmcm$-H$_2$I stretch slightly to 0.817~\AA{} and the distance between them shrinks to 1.198~\AA{}. This suggests that if $Cmcm$-H$_2$I is stable at higher pressures, these chains may polymerize resulting in a 1-D hydrogenic motif, as has been predicted for $R\bar{3}m$-SrH$_6$ by 250~GPa \cite{Zurek:2013f}. 

A $P6/mmm$ symmetry H$_4$I structure also comprised the 100~GPa convex hull, and it was found to be the most stable H$_4$I phase between 90-200 GPa. Phonon calculations revealed dynamic stability at 100 and 150~GPa. This phase is composed of 1-D chains of iodine atoms with an I-I distance of 2.812~\AA{} and H$_2$ molecules measuring 0.799~\AA{}, both lying parallel to the $c$-axis as illustrated in Fig.\ \ref{fig:H2Ifigure1}(b). The intermolecular distances between the dihydrogen molecules (1.875-2.013~\AA{}) and between H-I ($\sim$2.13~\AA{}) are too long for bonding interactions. Similar to H$_2$I, the iodine donates electrons (the Bader charges are (H$^{-0.05}$)$_4$I$^{0.21}$), and the ELF indicates high electron localization only within the dihydrogen molecules. At 200~GPa the H$_2$ units are not yet close enough to polymerize. 

The total and projected DOS plots for $Cmcm$-H$_2$I and $P6/mmm$-H$_4$I at 100~GPa resemble each other, see Fig.\ \ref{fig:H2Ifigure1}, which is not surprising considering their structural similarities.  Both phases are good metals with the DOS at $E_F$ being 0.061 and 0.049~eV$^{-1}$ states/(valence electron), respectively. The DOS can be split up into 3 regions: $<$-13~eV, -13 to -5~eV and above -5~eV it is primarily of I $s$, H$_2$ $s$ and I $p$-character, respectively, but the H$_2$ $s$-states have a large degree of overlap with both iodine $s$ and $p$. The metallicity is primarily due to the iodine $p$-states, but for $P\ge$100~GPa the hydrogenic states also provide a substantial contribution to the DOS at $E_F$.

Pressure causes the intramolecular I-I bond in $Cmca$-I$_2$ to stretch, and the intermolecular distances to decrease. At $\sim$16~GPa this phase undergoes an insulator to metal transition \cite{Drickamer:1963a}, which is superseded by an incommensurate phase at 23~GPa and the formation of a monoatomic phase within which the I-I distances are ``equalized'' by 30~GPa \cite{Kenichi:2003a}. Experiments yielded a bond length of 2.75~\AA{} at ambient conditions, whereas the nearest neighbor bonds in the fully dissociated monoatomic crystal measured 2.89~\AA{} \cite{Kenichi:2003a}. In our calculations bond equalization occurs near 35~GPa with I-I distances of 2.98~\AA{}. Interestingly, we find that $Cmcm$-H$_2$I and $P6/mmm$-H$_4$I are already metallic at 1~atm, and the DOS at $E_F$ exhibits primarily iodine $p$-character. The metallicity is a result of the monoatomic iodine present in these phases at ambient conditions, wherein the nearest-neighbor distances are found along 1-D chains measuring 2.93~\AA{}.

Experimental measurements yield a $T_c$ of 1.2~K at 28~GPa for iodine \cite{Shimizu:1994a}. Theory has shown that under hydrostatic conditions $T_c$ decreases with increasing pressure, falling to $<$1~K above 100~GPa \cite{Shirai:1996a,Zou:2009a}. Our calculations for the $Fm\bar{3}m$ phase of iodine using the Allen-Dynes modified McMillan equation \cite{Allen:1975a} yield $T_c$ values that are in agreement with these studies, see Table \ref{tab:Tc}. They further confirm that the logarithmic average frequency, $\omega_\text{log}$, which is low due to the heavy mass of iodine, increases slightly with pressure \cite{Zou:2009a}. But the decrease in the electron-phonon coupling, $\lambda$, with increasing pressure dominates, leading to a drop in $T_c$ at higher pressures. 

\begin{figure}
\begin{center}
\includegraphics[width=\columnwidth]{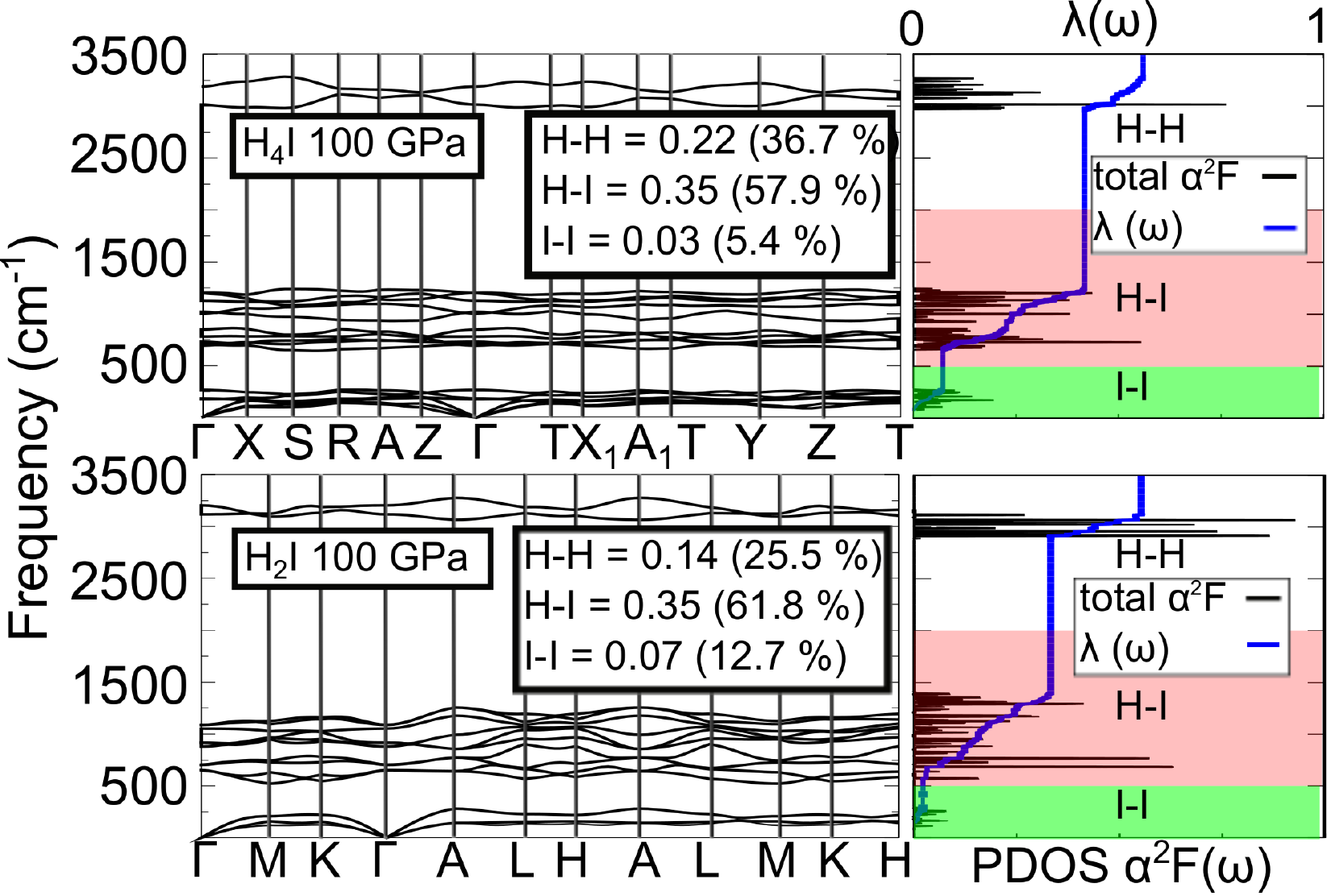}
\end{center}
\caption{Phonon band structure, the Eliashberg spectral function, $\alpha^2F(\omega)$, and the electron-phonon integral, $\lambda(\omega)$, for $P6/mmm$-H$_4$I and $Cmcm$-H$_2$I at 100~GPa. The insets provide the contribution to $\lambda$ (absolute value and percent) from vibrational modes that are comprised of iodine vibrations (I-I), motions of hydrogen and iodine atoms (H-I), and the H$_2$ vibron (H-H).}
\label{fig:Tc}
\end{figure}

What effect would incorporation of H$_2$ within lattices consisting of monoatomic iodine, in the geometries adopted by H$_2$I and H$_4$I, have on  $\omega_\text{log}$, $\lambda$ and $T_c$? At 100~GPa we find that $\omega_\text{log}$ becomes larger as the H:I ratio increases, see Table \ref{tab:Tc}, because of the light mass of hydrogen. In addition, the $\lambda$ of H$_4$I is slightly larger than that of H$_2$I, and these are both significantly larger than that of pure iodine. The most significant contribution towards $\lambda$ in both of the iodine polyhydrides stems from vibrational modes between 450-1300~cm$^{-1}$ that involve both hydrogen and iodine motions, with the hydrogen vibron also contributing substantially towards $\lambda$, see Fig.\ \ref{fig:Tc}. Vibrations that involve only the iodine atoms do not contribute much to the total electron phonon coupling. The computed $T_c$ of the iodine polyhydrides for $\mu^*=0.1$, 7.8 and 17.5~K for H$_2$I and H$_4$I at 100~GPa, is significantly larger than that of monoatomic iodine because H$_2$ increases both $\omega_\text{log}$ and $\lambda$. 

We also examined the effect of pressure on the $T_c$ of H$_4$I. At 150~GPa the $T_c$ was computed to increase to 20.4~K as a result of the larger $\omega_\text{log}$, which is due to the larger frequency range of the iodine and mixed iodine-hydrogen vibrational modes. In contrast, the frequency associated with the H$_2$ vibron decreased slightly with increasing pressure as a result of bond weakening within the hydrogen molecule, which measured 0.794 and 0.806~\AA{} at 100 and 150~GPa, respectively. A Bader analysis revealed that the softening of the hydrogen vibron can be attributed at least in part to the increased electron donation from iodine to H$_2^{\delta-}$ at higher pressures, which lengthens this bond. The slight decrease in the electron-phonon coupling with increasing pressure stems primarily from the smaller values of $\lambda$ associated with modes arising from motions involving iodine atoms. In contrast, the $\lambda$ associated with the H$_2$ stretch increased slightly under pressure.

\begin{table}
    \centering
    \caption{Electron-phonon coupling parameter ($\lambda$), logarithmic average of phonon frequencies ($\omega_{\text{log}}$) and estimated superconducting critical temperature ($T_c$) for values of the Coulomb pseudopotential ($\mu^*$) of 0.1 and 0.13 for $Cmcm$-H$_2$I, $P6/mmm$-H$_4$I and $Fm\bar{3}m$ iodine at various pressures.}
        \begin{tabular}{c c c c c c c c}
\hline
System & Pressure (GPa) & $\lambda$ & $\omega_{\text{log}}$ (K)  & T$_c^{\mu^*=~0.1}$ (K) & T$_c^{\mu^*=~0.13}$ (K) \\
\hline
H$_2$I & 100 & 0.51 & 608.7 &    7.8 & 4.8 \\

H$_4$I & 100 & 0.58 & 854.3 &    17.5 & 12.0\\

H$_4$I & 150 & 0.56 & 1097.4 &   20.4 & 13.7\\

I  & 100 & 0.31 & 251.6 &     0.16 & 0.03\\

I  & 150 & 0.25 & 304.5 &     0.02 & 0.00\\
\hline
\end{tabular}
\label{tab:Tc}
\end{table}

In conclusion, evolutionary searches have identified three hitherto unknown iodine polyhydride phases with unique stoichiometries that become thermodynamically and dynamically stable above 30~GPa. $P_1$-H$_5$I, composed of zigzag [(HI)$^{\delta+}]_\infty$ chains and H$_2^{\delta-}$ molecules, is found on the 50~GPa convex hull and has the lowest $\Delta H_f$ of any phase examined until 90~GPa. Metalization occurs as a result of pressure induced broadening of the iodine $p$ and hydrogen $s$-states, at pressures likely outside of the stability range of this structure. The DOS at $E_F$ suggests that the $T_c$ of this phase is likely to be low. $Cmcm$-H$_2$I and $P6/mmm$-H$_4$I both lie on the 100, 150 and 200~GPa convex hulls. They are comprised of monoatomic iodine lattices that render these phases metallic by 1~atm, and H$_2$ molecules. A Bader analysis suggests a slight transfer of charge from iodine to H$_2$.  The metallicity persists as H$_2$I and H$_4$I become dynamically and thermodynamically stable phases. The DOS at $E_F$ is primarily due to the iodine $p$-states, with an admixture of hydrogen $s$. At 100~GPa the $T_c$ of H$_2$I and H$_4$I is estimated as being 7.8 and 17.5~K, respectively. The presence of H$_2$ significantly enhances $T_c$ in comparison to pure iodine because the light atomic mass of hydrogen increases $\omega_\text{log}$, and because vibrations involving H and I atoms, as well as the H$_2$ vibron, contribute substantially to $\lambda$. As the quest for high-temperature superconductivity in hydrides gains momentum, we look forward to the eventual synthesis of these, and other intriguing phases.

We acknowledge the NSF (DMR-1505817) for financial, and the Center for Computational Research (CCR) at SUNY Buffalo for computational support. A.S.\ acknowledges financial support from the Department of Energy National Nuclear Security Administration under Award Number DE-NA0002006, and  E.Z.\ thanks the Alfred P. Sloan Foundation for a research fellowship (2013-2015). We thank Duck Young Kim for fruitful discussions.


\end{document}